# 2D material-based mode confinement engineering in electro-optic modulators

Zhizhen Ma[1], Behrouz Movahhed Nouri[1,2], Mohammad Tahersima[1], Sikandar Khan[1], Hamed Dalir[1], Volker J. Sorger[1]

*Abstract*— The ability to modulate light using 2-dimensional (2D) materials is fundamentally challenged by their small optical cross-section leading to miniscule modal confinements in diffraction-limited photonics despite intrinsically high electro-optic absorption modulation (EAM) potential given by their strong exciton binding energies. However the inherent polarization anisotropy in 2D-materials and device tradeoffs lead to additional requirements with respect to electric field directions and modal confinement. A detailed relationship between modal confinement factor and obtainable modulation strength including definitions on bounding limits are outstanding. Here we show that the modal confinement factor is a key parameter determining both the modulation strength and the modulator extinction ratio-to-insertion loss metric. We show that the modal confinement and hence the modulation strength of a single-layer modulated 2D material in a plasmonically confined mode is able to improve by more than 10x compared to diffraction-limited modes. Combined with the strong-index modulation of graphene the modulation strength can be more than 2-orders of magnitude higher compared to Silicon-based EAMs. Furthermore modal confinement was found to be synergistic with performance optimization via enhanced light-matter-interactions. These results show that there is room for scaling 2D material EAMs with respect to modal engineering towards realizing synergistic designs leading to high-performance modulators.

## I. INTRODUCTION

With device down-scaling the capacitive delay of electronics increasingly limits communication overhead [1]. Photonics on the other hand bears a notion of parallelism since a) photons only interact through media with each other, and b) the bosonic character allow a state to occupy the same quantum state. As a result, optics is therefore one possible candidate-of-choice for data communication. While the prospects for photonics are known, optoelectronic devices need either be footprint-density competitive to electronics or outperform electronics, or both, in order to enter the chip market. Loosely put, the expected break-even performance of diffraction limited opto-electronic devices relative to electronics has to be about 100 times to make up for the larger footprint (electronics = 20 nm, photonics diffraction limit = λ/2n ~ 200 nm, where λ = telecom wavelength). This barrier to entry could be lowered by scaling-down opto-electronics beyond the diffraction limit, which is actually synergistic to device performance discussed exemplary one photonic device here.

The device performing the electrical-optical conversion is the electro-optic modulator (EOM), which is an integral building block in photonic integration. The device performance of the EOM greatly depends on a variety of interrelated design concepts; a) the underlying waveguide determining coupling and propagation losses, b) the overlap fraction of the active material with the optical mode, c) the strength of the optical index change being altered [2,3], and d) subsequent impacts on energy efficiency, modulation speed, footprint, and optical power penalty [4-8]. Regarding the latter, we note that both the energy efficiency and modulation 3-dB speed bandwidth depend on both the electrical and optical limitations, with details given in ref [9]. Previous work focused on addressing a) through d) in an add-hoc manner [2, 10-12]. While using optical cavities allow enhancing the interaction strength of light with the active material, such approaches can limit the energy-per-bit efficiency, which scales proportional with device volume [9]. Here we show that high EOM modulation efficiency performance (i.e. extinction ratio (ER) per unit length) can be achieved by addressing the modal confinement in combination with a highly index-changing 2D material [9]. Hence the choice of the modal field distribution becomes important to increase the optical density of states non-resonantly [12-14]. The latter connects to the understanding that the energy-per-bit of an EOM is inversely proportional to the Purcell factor, $F_p \sim Q/V_m$, where $Q$ is the quality factor of the system or cavity, and $V_m$ the optical mode volume [15]. We note that this scaling can be influenced by the selected modulation mechanism chosen. For instance, in phase-shifting-based electro-optic modulation the E/bit scales with $(F_pQ)^{-1}$ and not just $1/F_p$. However, while a), c) and d) were previously investigated for EOMs including emerging materials, the particular effect of the modal confinement b) of the modulation capability is still outstanding as explored here. Here the question arises whether 2D materials are even viable options for modulators due to their molecule crossection. This points to the need for a detailed analysis of

This work was supported in part by the U.S. Air Force Office of Scientific Research under Grant AF9550-14-1-0215 and AF9559-14-1-0378.
[1]Department of Electrical and Computer Engineering, The George Washington University, Washington, D.C. 20052, USA (e-mail: sorger@gwu.edu).
[2] Optelligence LLC, Upper Marlboro, MD, 20772, USA.

achievable modal confinements of the optical mode with a 2D material. Aside from this question, the rational to even consider 2D materials for optoelectronics and more specifically EOMs is threefold; i) these materials are synergistic to the high-$F_p$ form factor (small $V_m$) EOM designs, ii) are patternable enabling ease in semiconductor processing, iii) enable strong material index modulation [16,17]. Regarding the latter, the exciton binding energy (oscillator strength) is sensitive to the Coulomb potential between the electron-hole pair, which in turn can be screened by its dielectric environment. Thus, the higher the dielectric constant, the weaker the screened potential; as dimensionality reduces, the effective dielectric constant also reduce since the exciton now interacts more of the surrounding than the material itself.

There are a variety of 2D materials for electro-optic modulation that have proposed or shown functionality. Here we consider graphene, simply for its known properties, however other 2D materials could be considered as well. Graphene has shown electro-optic response via Pauli blocking in for near IR frequencies and modulating functionality [4, 18]. Indeed effort has been made in integrating graphene with plasmonics with the purpose of modulation [2, 14]. Yet the anisotropy of 2D films like graphene introduce challenges with respect to polarization alignment [19, 20]. As a result, plasmonic approaches, thus far, have shown low modulation capability and non-synergistic device designs despite graphene's strong index modulation potential [14]. Nevertheless, graphene phase modulation shows tens of GHz fast modulation, however relies on the strong feedback from a mirroring cavity leading to non-compact footprints and temperature sensitivities [21].

Here we investigate the effect of the modal confinement onto modulation capability; we compare different modal designs by choosing graphene as an exemplary active 2D material. Taking polarization limitations of 2D materials into account for the optical mode, we show that only a limited design window exists. Testing such modal confinement engineering, we finish by showing results based on a plasmonic slot waveguide electro-optic absorption modulator (EAM) which is able to provide a very high modulation capability with a modal design that is synergistic to the 'right' (in-plane) electric field component of graphene.

## II. UNIQUE ELECTRO-OPTIC PROPERTY OF GRAPHENE

Graphene shows great anisotropic material properties given its dimensions: in its honeycomb like lattice plane, the in-plane permittivity ($\varepsilon_{||}$) can be tuned by varying its chemical potential $\mu_c$, whereas the out-of-plane permittivity is reported to remain constant around 2.5 [22]. In this work, graphene is modeled by a surface conductivity model $\sigma(\omega, \mu_c, \Gamma, T)$, where $\omega$ is the radian frequency, $\mu_c$ is the chemical potential, $\Gamma$ is the scattering rate and $T$ is the temperature. The conductivity of graphene is given by the Kubo formula [23]:

$$\sigma(\omega, \mu_c, \Gamma, T) = \frac{je^2(\omega - j2\Gamma)}{\pi\hbar} \left[ \frac{1}{(\omega - j2\Gamma)^2} \int_0^\infty \varepsilon \left( \frac{\partial f_d(\varepsilon)}{\partial \varepsilon} - \frac{\partial f_d(-\varepsilon)}{\partial \varepsilon} \right) d\varepsilon - \int_0^\infty \frac{f_d(-\varepsilon) - f_d(\varepsilon)}{(\omega - j2\Gamma)^2 - 4(\varepsilon/\hbar)^2} d\varepsilon \right] \quad (1)$$

where $f_d(\varepsilon) = 1/\{\exp[(\varepsilon - \mu_c)/k_B T] + 1\}$ is the Fermi-Dirac distribution function, $\hbar$ is the reduced Planck's constant $\hbar = h/2\pi$, and $e$ is the electron charge. The chemical potential $\mu_c$ is a function of gate voltage $V_g$ and the capacitor model, which we explicitly studied in the device section below. Fig. 1 (a) shows the in-plane conductivity $\sigma_{||}$ calculated from Kubo formula at wavelength $\lambda = 1550$ nm. Then the in-plane permittivity is calculated by $\varepsilon_{||} = 1 - \frac{\sigma_{||}}{j\omega\varepsilon_0 \Delta}$, where $\Delta = 0.35$ nm is the thickness of graphene sheet, and the refractive index is calculated from the permittivity (Fig. 1 (b)).

Note that the drastic change in graphene's imaginary refractive index, $\kappa$, is due to the strong effect through Pauli blocking, thus making graphene a naturally suitable material for EAMs (Fig. 1 (b)). When a voltage applied to the graphene/oxide capacitor, carriers begin to accumulate on the graphene sheet and thus the chemical potential is tuned with the gate voltage. As $|\mu_c|$ reaches half of the equivalent photon energy ($1/2\ h\nu$), the interband transition is blocked hence the absorption greatly decrease (Fig. 1 (c)).

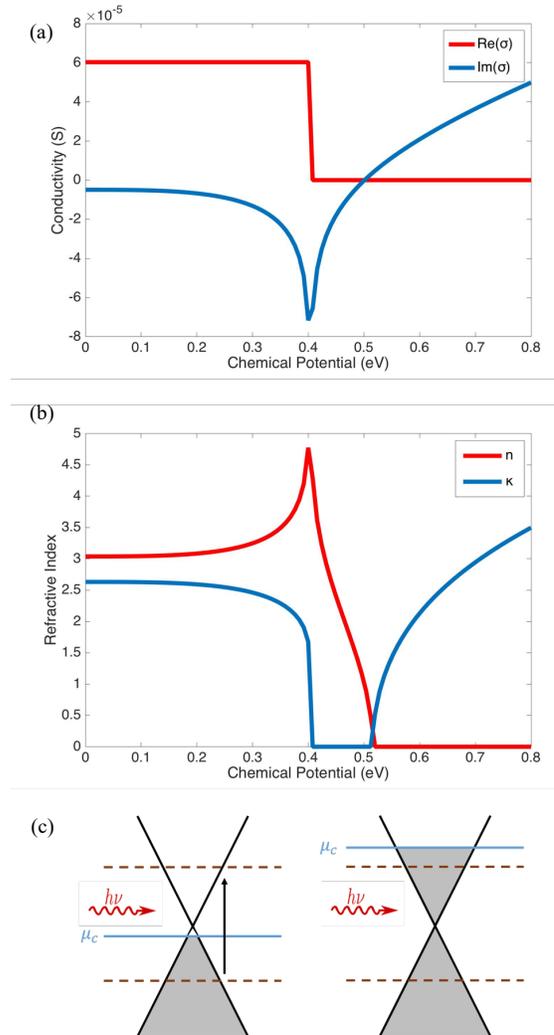

Fig. 1. Graphene modulation physics. (a) and (b) Graphene conductivity and refractive index as a function of chemical potential. (c) Pauli blocking effect in graphene: when the chemical potential is larger then $1/2h\nu$, the interband transition is blocked.

## III. GRAPHENE EAM PLATFORM STUDY

### A. Surface Plasmon Polariton for Graphene

Following the arguments from the introduction, the Purcell factor helps to improve the modulator performance such as the energy efficiency. Enhancements of the light-matter interaction (LMI) can be obtained non-resonantly to provide via modal confinement, by using the plasmonic waveguide for on-chip guided wave applications [14,24]. However, integrating plasmonic waveguides might not be straightforward, in fact, is actually incompatible unless specifically engineered; graphene is commonly placed at the interface between metal and dielectric material. Here, the surface plasmon polaritons (SPP) generated by the incoming photon and metal dielectric interface are predominantly out-of-plane (TM), and only a very small portion of the electric field is projected to graphene's in-plane direction (Fig. 2 (a)). This contradicts the aforementioned property of graphene requiring in-plane electric field components to actively tune its optical index. That is, vanishing index change is expected when the electric field applied, such as the case in modulation, is perpendicular relative to the graphene film. This anisotropy makes the integration of graphene and plasmonic waveguide fundamentally challenging. However, there are options to address this issue. Here we propose a solution to place graphene in the lateral plane with the SPP electric field penetrating direction by using plasmonic slot waveguide (Fig. 4). In this design the absorption of a single layer graphene would benefit from a maximum in-plane electric field component (Fig. 2 (b)), which results in a higher modulation performance compare to previous work reported [4]. The detailed device design and expected performance are discussed in the following section.

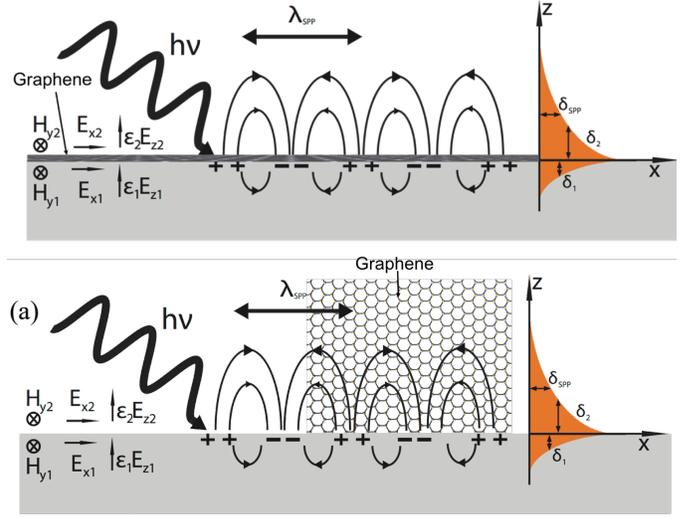

Fig. 2 SPP mode interaction with graphene. (a) A flat graphene placed on a metal-dielectric interface resulting in an out-of-plane does not interact strongly with the atom-thin graphene material due to the lack of polarization. (b) Graphene normal to interface is aligned to be in-plane with the SPP mode, and can be (in principle) used for electro-optic functionality.

### B. Platforms Study

2D materials show their unique electro-optic tunability when the electric field is in the lattice plane, but not perpendicular to it. Hence, here we propose a metric capturing the confinement factor to characterize the light-2D material interaction as:

$$\gamma_{2D} = \frac{\int_{2D} |E_{in}|^2 dS}{\int_S |E|^2 dS} \qquad (2)$$

and similarly for bulk material we have:

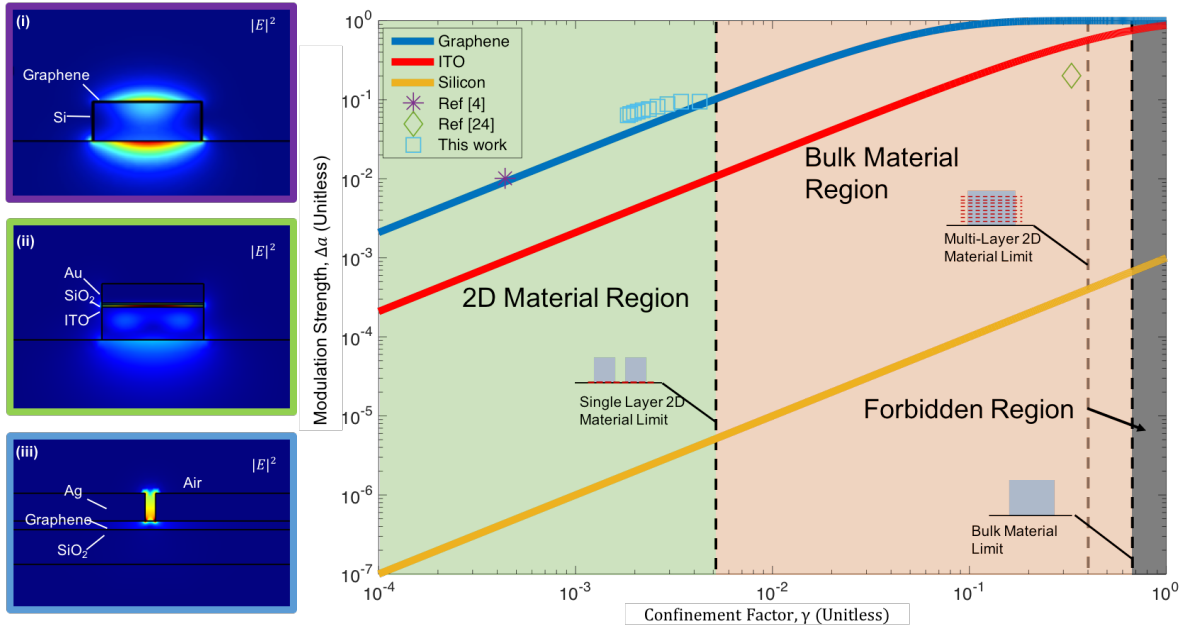

Fig. 3 Analytical result of modulation for three different materials is plotted here (solid lines). The confinement factor limits for single- and, multi-layer stack 2D material, and diffraction limit bulk material separate the figure into different regions. The single layer 2D material confinement factor is low for diffraction-limited modes (i), but can approach that of bulk material when engineered (iii). A forbidden region exists due to mode leakage. Three different EAM configurations (i-iii) were studied numerically, and data were plotted as data points here in the figure, which shows a good agreement with analytical result. See 'Supporting Information' for details on the optical permittivity's used. $\lambda$ = 1550 nm.

$$\gamma_b = \frac{\int_b |E|^2 dS}{\int_s |E|^2 dS}$$
(3)

where $E_{in}$ denotes the in-plane electric field, and $E$ is the electric field. To capture this component and towards quantifying the modal confinement factor, we introduce the concept of modulation strength, which is the actually changed optical absorption due to ON and OFF state change of the EAM, defined as

$$\Delta a = 1 - e^{-2k_0 Im(\Delta n_{eff})z} \quad (4)$$

where $n_{eff}$ is the effective mode index, $k_0 = \frac{2\pi}{\lambda}$ is the free space wave number, $z$ is the propagation distance along the device region. In the following discussion, we analytically and numerically studied three different EAM waveguide cross-sections and show that how the confinement factor impacts the modulation strength, and hence directly influences the EAM performance.

ER has always been a key factor to characterize the performance of EAMs, however, ER is a function of both the ON and OFF state transmission. In this work the comparison is between 3 type EO materials (graphene, Silicon, ITO) in different waveguide configurations (SPP slot, hybrid plasmon polariton (HPP) [24] and diffraction limited photonic mode [4]), the modal change loss and extra coupling loss between ON and OFF state is not straightforward to conclude, thus here we use modulation strength instead of ER to determine the modulation performance between the ON and OFF state. Also, since the effective mode index is also a function of both environment and active material, which can not be easily obtained analytically, here we take the $Im(n_{eff})$ change between ON and OFF state to be only attributed to the active material, and proportional to its confinement factor, thus Equ. (4) could be simplified to

$$\Delta a = 1 - e^{-2\gamma k_0 \Delta \kappa z} \quad (5)$$

The analytical solution of modulation strength for three different EO effect material-waveguide options gives an insight into the modulation strength performance through sweeping the modal confinement (Fig. 3). The refractive index modulation data of ITO and Silicon was taken from previous experimental work [2, 24, 25]. Note that here we normalized the modulation strength to per micrometer (z = nominal length = 1 μm). Silicon as a conventional EO material, which has shown Kerr, and carrier injection effect, has rather low modulation strength, due to the small refractive index change. Hence either very long linear or high-Q cavity based devices were explored to accumulate sufficient modulation [26, 27, 28]. ITO of the transparent conductive oxide family, is an emerging EO material which has unity order change in refractive index when under electrical gating [2, 29].

The modulation strength as a function of modal confinement factor spans three general regions defined by both materials and fundamental electromagnetics limits, namely, '2D Material', 'Bulk Material', and 'Forbidden' regions (Fig. 3). The forbidden region is shown here is defined by the ratio of the field sitting outside the waveguide versus inside, and is about 65% for typical SOI waveguides. Thus, even for bulk waveguides whose refractive index could be in principle modulated entirely, do not allow for 100% confinement factors.

Even if such a diffraction-limited waveguide is increased the evanescent tail introduces limits. 2D materials are naturally limited to regions of small confinement factors similar to quantum well. For instance, if a bulk waveguide is selected and a single 2D material is added, then the observed [4] (purple star in Fig. 3) and predicted ((i) Fig. 3) confinement value is about 0.04%. If the optical mode is reduced such as via a plasmonic slot waveguide this value increases by more than 10x to about 0.5% (left black trend line Fig. 3). We connect the latter value to the maximum confinement found for a tiny metallic slot waveguide-based modulator hereby termed quantum dot modulator (QDM). For such a structure both the metal height and metal-air-metal gap are ~20 nm (see Fig. 4 for geometric definitions). This value (20 nm) can be understood from two pictures both relating to the fact that metallic confinement beyond 20 nm is not possible: a) the skin depth of plasmons at telecomm wavelengths is about 20 nm of nanometer, and b) the Purcell factor reduces dramatically beyond 10's nm small plasmonic cavities (see 'Supplementary Information') [3]. Another conceptually trivial option to increase the confinement factor is to stack multiple 2D materials inside a bulk material to increase the absorption similar to multiple quantum well designs. Assuming a 1:1 spacer to 2D-material ratio, the achievable confinement can be estimated by halfing that of the bulk case, since the spacer consumes about half of the geometrical cross-section of the waveguide resulting in about 40% confinement (grey dashed line Fig. 3). However, the implementation for this structure is somewhat unrealistic, due to the inability to electrically contact the graphene layers without completely altering the optical mode. To circumvent this problematic issue, in Figure 4 below we propose and

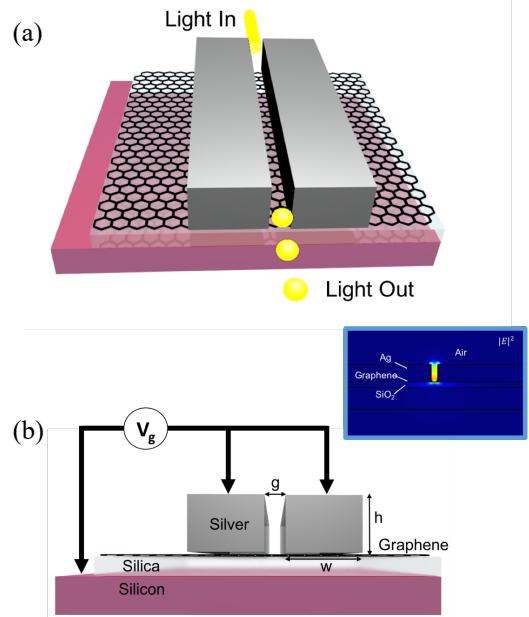

Fig. 4 (a) Device schematic of a plasmonic slot-waveguide EAM. (b) The silver slot provides plasmonic sub-wavelength confinement, and serves as a contact at the same time. The mode profile is illustrated as an inset. The Silicon substrate thickness is blow cut-off and functions as a back-gate for electrical gating graphene. Free parameters are the slot gap (g) and the metal height (h). See 'Supporting Information' for details of the material parameters.

investigate a EAM design deploying a plasmonic slot waveguide to provide sub-wavelength mode confinement, allowing for in-plane electric field of the active graphene layer.

Both the intrinsic material ability to change its index along with achieving a high confinement factor together determines the modulation strength. The relative strength is shown in the solid lines for graphene, ITO, and Silicon independent of the optical mode chosen as a function of nominal confinement factor (blue, red, yellow line Fig. 3). The reason for the observed higher modulation strength in graphene compared with other two bulk materials, ITO and Silicon, is the drastic refractive index change in graphene, resulting from Pauli blocking, where the transition from interband to intraband absorption blocks the electron-photon interaction in telecomm wavelength (blue vs. red line Fig. 3). However, ITO as a material follows the Drude model and the refractive index change due to its carrier accumulation is comparably lower. It is interesting to ask why ITO has a higher modulation strength than Silicon. In other words, would highly doped Silicon have the same index change as ITO? The answer is 'no' because the bandgap of Silicon is smaller compared to ITO. Thus, for the same doping level the amount of degenerate carriers in the conduction band is lower for Silicon, thus the contribution of those carriers to the index modulation is (relatively speaking) lower, explaining the lower Silicon modulation strength compared to ITO (yellow line Fig. 3).

Investigating the particular device designs per example reveals insights in how close-to-optimal a particular modal waveguide design is to these idealize solid trend lines of Figure 3. To understand this deeper, we numerically studied three specific design of EAM designs; a diffraction limited Silicon waveguide with graphene placed on top [4], a HPP ITO modulator [24], and SPP slot plasmonic design investigated here. Finite element method (FEM) and finite-difference time-domain (FDTD) simulation was used for all three designs, where the mode profile from eigenmode analysis is shown in insets of Figure 3, and the results of confinement factor and modulation strength is plotted as data points.

One can see that for diffraction limited mode graphene modulator, the confinement factor is only 0.04%, which corresponds to a modulation strength of 2%, which is consistent with reported experimental result [4]. This contrasts an increase by 10x to about 0.4% for the slot plasmonic design for the smallest metal height and gap width (see 'Supplementary Information'). This value is interestingly close to the expected limit for the plasmon QDM designs for single layer 2D materials (gap width = 30, slot height = 20 nm). The 0.4% confinement factor at this point relates to a relatively high modulation strength of 10%. It is interesting to note that this value is just 1/3$^{rd}$ of that of the HPP ITO modulator with a 30x thicker active material, and highlights the ability of engineering the confinement factor with 2D material to approach modulation performance similar to that of traditional bulk materials. In fact if now, multi-layers are deployed, then even a few 2D layers would outperform bulk materials, yet more details are necessary to quantitatively support this argument. We note that the discrepancy between analytical solution and numerical solution in Figure 3 is likely due to the fact that in the analytical solution we ruled out the effect of the overall effective index change, but used Equ. (5) instead. However in the numerical solution the modulation strength is a function of effective index but not only of the confinement factor and refractive index change of active material.

As an interim conclusion we see that the confinement factor is a key parameter for the EAM performance and an indicator for the LMI not only graphene but also other 2D materials, and one could therefore estimate the device performance for multi functional optoelectronic devices such as light sources, routing switches, or photodetectors by performing an confinement factor study. Indeed we find that the due to the limitation of the physical dimension of 2D material, the confinement factor can never be higher when compare to a bulk material. However with appropriate modal engineering the overall device performance potentially surpass that of bulk design be even better. The latter might especially true given the scaling arguments from the introduction [9]. To test this hypothesis we further explore device performances of a graphene-based plasmonic slot waveguide EAM.

## IV. Device design and performance

The device design is depicted in Fig. (4), it consists of a Silicon substrate which has a thickness below cut-off to ensure the light is propagating only alone the plasmonic slot, which at the same time serves as a back gate for capacitive gating graphene. A 25 nm Silica oxide layer atop the Silicon substrate, was previously found for a plasmonic slot waveguide to provide adequate mode confinement [30]. Key to this design is that the silver slot here (i) provides a sub-wavelength plasmonic confinement which would enhance the LMI with graphene, (ii) allows the SPP mode to propagate in-plane with graphene sheet, increasing the aforementioned confinement factor, (iii) provides a metal contact which serves as an electrical electrode and heat sink at the same time.

### A. Performance Optimization

Device dimensions used for optimizing the performance include the air gap width (g), silver slot height (h) and silver slot width (w) as labeled in Fig. (4). Commercial FDTD software (Lumerical FDTD) was used to sweep both air gap from 30 nm to 200 nm and slot height from 20 nm to 220 nm. Here we track the metric ER per nominal device length (here per micrometer). We find that the ER could reach as high as 1.2 dB/μm when the gap is 30 nm and slot height only 20 nm, which is more than 13x times higher compared to single layer graphene EAM work [4] (Fig. 5(a)). We note that even in the low ER region, the ER is still higher than 0.2 dB/μm, which to the best of our knowledge exceeding all demonstrated modulation performance using a single layer graphene. The advantage of using a plasmonic slot waveguide is that unlike Silicon diffraction limited, the plasmonic mode has no cut-off dimension, which means the physical dimension of the device could be about 4x narrower compared to diffraction limited devices (2 x 25 nm = skin depth of metal + gap = 10 nm results in 60 nm total device width).

However, a particular design challenge is minimizing the insertion loss of the device induced from the silver ohmic loss and potential coupling loss to the small mode. Increasing the air gap would reduce the imaginary part of the effective index, which would benefit the ON state transmission, at the same time, increasing the slot height reduces the energy leakage to the environment, but both decrease the ER. Thus the optimization of ER and IL in the graphene slot EAM is a trade off. We deploy the figure of merit (FOM) for this design as $FoM = \frac{ER}{IL}$ (both in dB/µm) (Fig. 5(b)) [2]. The results indicate that when the air gap is minimized to 30 nm and the slot height is blow 100 nm, the overall device performance could reach an optimized point maximizing at 0.75. Note that in this work, we are just designing the single layer graphene device to illustrate how the confinement factor of 2D material could be enhanced by slot plasmonic mode, hence achieve the optimized device performance. One could always place a second layer graphene on top or also underneath the device to double the modulation effect [31], which we expect would be a breaking nature high extinction ratio even better than traditional bulk EO material.

### B. Energy Consumption and Speed Consideration

The electrical energy consumption of the device could be evaluated by capacitive dissipation via $\frac{1}{2}CV^2$, where the gating voltage is calculated from

$$|\mu_c| = \hbar v_F \sqrt{\pi a_0 |V_g - V_{Dirac}|} \quad (6)$$

where $v_F = 10^6 m/s$ is the Fermi velocity of graphene, $a_0 = \varepsilon_r \varepsilon_0 / de$ is yielded from the simple capacitor model. $V_{Dirac}$ denotes the initial doping level of graphene, which is a finite number here ignored. The bias voltage is calculated as $\pm 2.5 V_{pp}$, then the energy consumption is a function of the capacitance of the whole device. Here we consider a 3 µm long device, the width of the metal slot is calculated from the skin depth of silver, to avoid mode leakage from metal, 5 times of skin depth is usually chosen to be the metal thickness. By using $\delta_m = \frac{1}{k_0}|\frac{\varepsilon_m' + \varepsilon_d}{\varepsilon_m'^2}|^{\frac{1}{2}}$ [32], the skin depth into silver is around 20 nm, thus the slot width is chosen to be 100 nm. The energy consumption is equal to 400 aJ/bit. From eqn. (6) and the capacitive energy we see that the E/bit can be further lowered by improving the electrostatics such as by introducing high-k dielectrics or reducing the gate oxide thickness, or reducing the drive voltage. Naturally, the latter will result in a reduced modulator performance (i.e. extinction-ratio), which could impact the interconnect performance. From Figure 5(a) we find that for small slot-heights and air-gaps (around 20nm each) an ER of 1.2dB/µm can be achieved. This means that for a short-length interconnect [2,33] requiring only a 3dB modulation a 2.5 µm compact device is needed, resulting in a 320aJ/bit efficient modulator. However, the above analysis was done for a single-sheet of graphene. Implementing a dual-modulation scheme [34] would half the energy consumption to about 160aJ/bit.

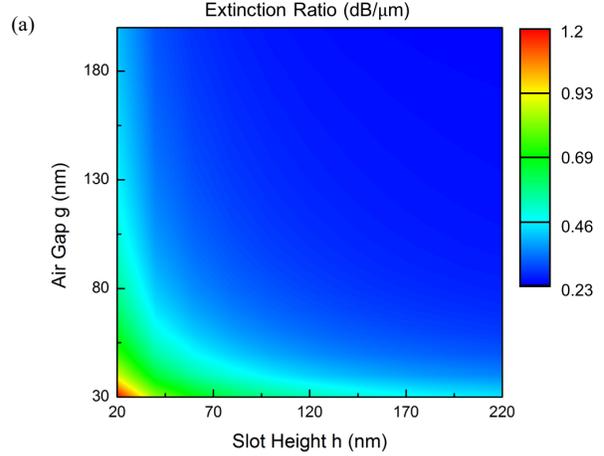

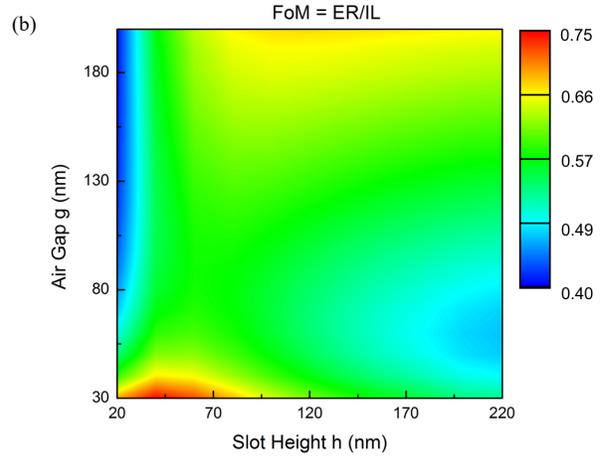

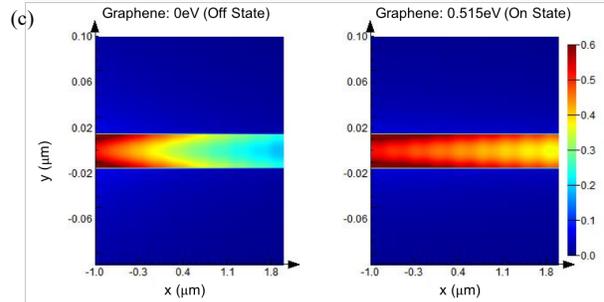

Fig. 5 The extinction ratio and ON state transmission of the device. (a) ER greatly improves as both gap and height decrease reaching ER = 1.2 dB/µm. (b) The ER/IL plot shows that the optimized dimension of the device is gap = 30 nm and slot height below 100 nm. (c) ON and OFF state power flow distribution from 3D FDTD simulation. λ = 1550 nm.

The bandwidth of the modulator can be estimated by $1/(2\pi RC)$, which is constrained by the values of capacitance and resistance. Making a simple estimation using these values however might not be insightful as this approach yields speeds in the sub-picosecond range. Such short time scales are somewhat unrealistic for electrical drivers and average power dissipation might be relatively high as well. In addition other carrier and mobility related effects are likely to be limiting factors and a proper transient analysis should be conducted in future work. A device consideration would be to use Silicon as a back gate; here the resistance of Silicon is rather high compared to metal, however, since there is no light propagating in Silicon substrate, it could be selectively doped to a very high carrier concentration to achieve low resistivity.

*C. Coupling Method*

For a photonic on-chip network device, it is essential to have a good coupling efficiency to the network or free space. For coupling to free space, one could adopt the plasmonic antenna, which has a 10% coupling efficiency for a 100 nm height and 30 nm gap [35-37]. To couple to other on-chip waveguide modes, a recently reported adiabatic taper has experimentally showed that a coupling efficiency of 65% could be achieved [38]. However, with the advent of functional and small-footprint EAMs, one could consider all-plasmonic on-chip networks for short range interconnects, where the plasmon detection could be conducted via hot electron detection.

*D. Narrow Slot-Device Prototyping*

A main challenge of the slot-enhanced graphene detector is the realization of a narrow metallic slot. Fortunately, the simulation results show best performance also for a shallow slot depth (Fig. 5a, lower left corner), hence the aspect ratio for the nanofabrication are 1:1, thus elevating the task. Nonetheless a sub 20nm precise metal gap is challenging for the one side due to the tight alignment requirements between each slot 'wall' and secondly due to the grain-boundary size of typical metal deposition leading to poly crystalline nanostructures. The latter can be alleviated via thermal annealing techniques. Alternatively, ALD-based metal deposition processes (or MOCVD processes) could be developed over PVD methods to improve (reduce) grain boundaries. Estimating an ultimate limit, we arrive around 5nm gap sizes, again, due to the granularity of the metal.

In our recent work [35] we explored and realized metal slots down to 11nm in with deploying a 2-EBL-step approach, where each sidewall is written separately and stepped in 5nm increments. Resulting slots with a 2D material monolithically integrated underneath the slot are shown in Figure 6, with 6a showing a conceptual implementation of the overall structure and slot realization (Fig. 6b). Interestingly, the smaller the slot gap is, the more light is seen by the 2D material rather than parasitic metal slot (Fig. 6c). Note, the 2D material layer could also be a heterogeneous multi-layer, such as a dual-graphene layer in a push-pull configuration, as discussed above and in ref [30-41]

## V. DISCUSSION

In general, it is worth asking what fundamental performance scaling vectors exist for 2D-based modulators, or thin-film based modulators [9]. Performing a capacitive-based energy-per-bit analysis shows that the performance of the modulator scales with the device volume (Fig. 7, top). Hence, a smaller mode-volume offers higher modulation efficiency. Such nano-optical scaling effect makes therefore a strong case for thin-film based EOMs such as those based on 2D materials. Beyond formfactor scaling which offers a high and synergistic mode-overlap with a 'squeezed' optical mode, 2D materials offer further advantages for monolithic integrated photonic circuits to include lithographically patternable, they do not require any substrate matching (unlike III-V materials) and therefore provide a quantum-well to be placed on any substrate. They further offer performance related physics, such as a high index and index-modulation potential due to the low dimensionality and associated coulomb screening effects.

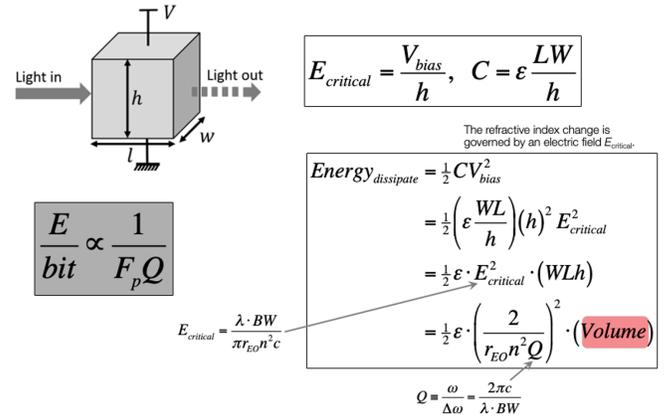

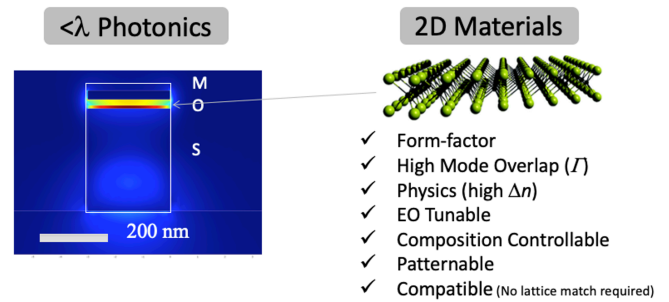

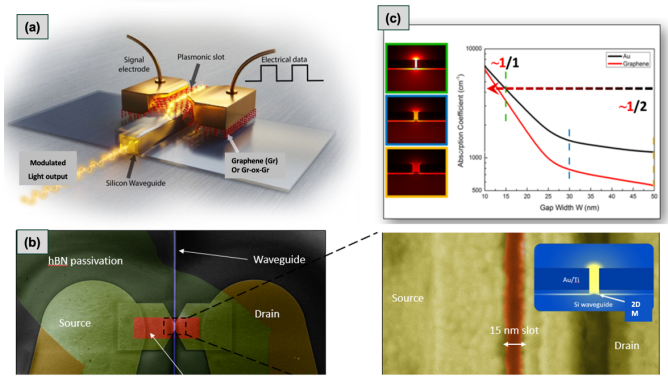

Fig. 6 Prototyping of a metal slot EOM integrated into a PIC waveguide platform. (a) Schematic of the overall design; a 2D material is sandwiched underneath two metal pads that are in close (10's nm) proximity. (b) Experimental photo. Inset: false-color SEM of the metal slot. Inset: eigenmode simulation. (c) Shrinking the metal slot doubles the amount of light seen by the 2D material, hence improves performance.

Fig. 7 (top) An analysis on the fundamental physical performance scaling vectors of nanostructured EOMs shows that the energy-per-bit of modulators scales with the inverse of the Purcell factor or proportional to the optical mode volume divided by cavity factor squared. (bottom) This small mode volume offers unique opportunities for 2D material-based E/bit-efficient devices. Beyond device performance, the case for 2D material-based devices rests in said small formfactor, which is synergistic to the mode-overlap factor, but also includes high index tunability, composition controllability, and being benign to the substrate, provided transfer options exist [42,43].

Next, we briefly discuss recent results of an 2D material-based and microring resonator (MRR)-enhanced modulator [44]. This device is based on a TMDC (MoTe$_2$) placed atop a Silicon photonic MRR ring, and is laterally electro-statically gated. The latter is key, since the gate oxide in this device is a lateral (horizontal) air gap of 1 μm (Fig. 8). Remarkably, we obtain a modulation of about 2dB with a very low (about 0.7dB) insertion loss (Fig. 8b). The 2D material is paced atop using our 2D material printer setup [42,43] offering wet-chemistry-free and cross-contamination free transfer of 2D material films or flakes. A small oxide bridges the Si waveguide and the TMDC (Fig. 8c). The low insertion loss can be explained by the material off-resonance with respect to the material exciton resonance line (Fig. 8d), hence the extinction coefficient (κ) is low. The eigenmode of the device can be optimized to maximize the field overlap with the 2D material sitting atop such as via adjusting the waveguide height, and/or width, or the spacer oxide between the waveguide and the TMDC. Since electrostatics scales with length, reducing the gate oxide from this laterally-gated scheme from 1 μm to 10 nm would reduce the drive voltage from currently 4V to 40mV, which in-turn would make this a 10's aJ/bit sensitive device. More work is needed to verify this in addition to speed and reliability measurements.

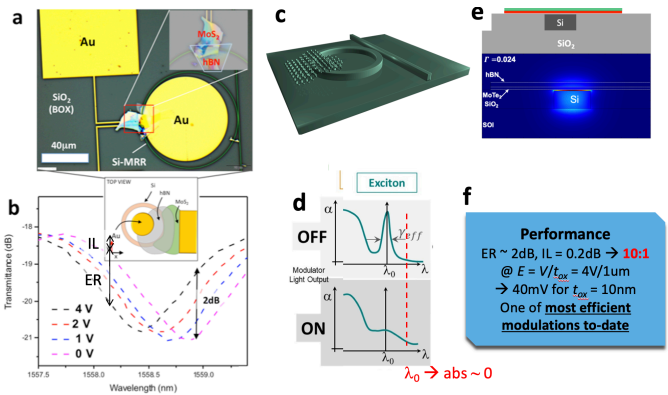

Fig. 8 Initial results of a 2D-material based MRR-enhanced modulator (a) fabricated device top-view. (b) DC modulation performance showing ER= 2dB with a low insertion loss of about 0.7dB. Inset: detailing the lateral gating scheme where a 1 μm wide air-gap is used as a gate oxide. (d) This TMDC (MoTe$_2$) is operated away from the exciton line on purpose to minimize the extinction coefficient and optimizing IL. (e) Cross-sectional schematic of the device and eigenmode showing a decent overlap with the actively modulated TMDC. (f) Performance summary of the EOM. Note, that the 4Vpp for a gate-oxide of 1000nm would result in a 40mVpp for a 10nm gate which in-turn would correspond to an E/bit of 10's aJ/bit, which is near the fundamental limit of a noisy channel.

Comparing this performance with other emerging material modulators such as those based on BTO or ITO, we find that 2D material modulators offer a very low insertion loss [45-48]. However, the obtainable ER might be limited by carrier saturation effects inside the thin 2D material, and hence may lead to only several dB of modulation unless interferometers are (e.g. MZI) are deployed, which could increase the device footprint significantly on-chip. Regarding modulation bandwidth (speed) of emerging material-based EOMs, modeling results have shown >100GHz modulation capability utilizing experimentally verified parameters such as mobility and resistivity for ITO-based devices, which have a much lower mobility (50-100 cm$^2$/Vs) as compared to graphene [49]. We like to point out that the concepts for achieving compact and efficient EOMs can also be deployed for high gain-bandwidth (BW) detectors; for example, deploying the above-mentioned slot and 2D material option has been predicted to break the typical gain-BW-product seen by detectors by making the device independent of transit-time via deployment of a slot structure [49].

## VI. CONCLUSION

In summary, we have studied the modal confinement factor of an electro-optic material and its impact on the modulation strength for electro-optic absorption modulation. We show that the modulation strength depends on both the material choice and the modal confinement factor. We find that if mode-engineering concepts are applied, the modulation strength of single-layer 2D materials can approach that of bulk materials. This is possible due to both the intrinsic strong index modulation potential of 2D materials originating from lack of screening the vertical electric field component stemming from its atomic thickness, and light-matter-interaction enhancements from waveguide mode engineering. This concept was found to be accurate when tested on a graphene-based plasmonic slot waveguide EAM design. The device shows confinement factors that are 10x higher compared to graphene designs using diffraction limited modes, resulting in modulation strengths of about 10%, an improvement of 10x (10$^3$x) over previous graphene (Silicon)-based EAM designs. High extinction ratio's of 1.2dB/μm were found from a single layer graphene EAM with proper mode-confinement engineering. The ability to outperform conventional bulk material device while delivering strong modulation and remaining within wavelength footprints and sub fJ/bit energy consumption shows a path for to advance on-chip optical interconnects, network-on-chips, or even for ASICs such as for machine-learning acceleration, such as photonic tensor core processors [50-54].

REFERENCES

[1] M. Waldrop, "The chips are down for Moore's law", *Nature*, vol. 530, no. 7589, pp. 144-147, 2016.
[2] C. Ye, S. Khan, Z. R. Li, E. Simsek, & V. J. Sorger, "λ-size ITO and graphene-based electro-optic modulators on SOI", *Selected Topics in Quantum Electronics, IEEE Journal of*, 20(4), 40-49, 2014.
[3] S. Sun, A. Badawy, V. Narayana, T. El-Ghazawi and V. Sorger, "The Case for Hybrid Photonic Plasmonic Interconnects (HyPPIs): Low-Latency Energy-and-Area-Efficient On-Chip Interconnects", *IEEE Photonics J.*, vol. 7, no. 6, pp. 1-14, 2015.
[4] M. Liu, X. Yin, E. Ulin-Avila, B. Geng, T. Zentgraf, L. Ju, F. Wang and X. Zhang, "A graphene-based broadband optical modulator", *Nature*, vol. 474, no. 7349, pp. 64-67, 2011.
[5] Y. Ye, Z. Wong, X. Lu, X. Ni, H. Zhu, X. Chen, Y. Wang and X. Zhang, "Monolayer excitonic laser", *Nature Photonics*, vol. 9, no. 11, pp. 733-737, 2015.
[6] M. Tahersima and V. Sorger, "Enhanced photon absorption in spiral nanostructured solar cells using layered 2D materials", *Nanotechnology*, vol. 26, no. 34, p. 344005, 2015.
[7] X. Gan, R. Shiue, Y. Gao, I. Meric, T. Heinz, K. Shepard, J. Hone, S. Assefa and D. Englund, "Chip-integrated ultrafast graphene photodetector with high responsivity", *Nature Photonics*, vol. 7, no. 11, pp. 883-887, 2013.
[8] G. Li, A. Krishnamoorthy, I. Shubin, J. Yao, Y. Luo, H. Thacker, X. Zheng, K. Raj and J. Cunningham, "Ring Resonator Modulators in

Silicon for Interchip Photonic Links", *IEEE J. Select. Topics Quantum Electron.*, vol. 19, no. 6, pp. 95-113, 2013.
[9] K. Liu, A. Majumdar, S. Sun and V. Sorger, "Fundamental Scaling Laws in Nanophotonics", Sci Rep. 6, 1, 37419 (2016).
[10] C. Ye, K. Liu, R. Soref and V. Sorger, "A compact plasmonic MOS-based 2×2 electro-optic switch", *Nanophotonics*, vol. 4, no. 1, 2015.
[11] K. Liu, C. Ye, S. Khan and V. Sorger, "Review and perspective on ultrafast wavelength-size electro-optic modulators", *Laser & Photonics Reviews*, vol. 9, no. 2, pp. 172-194, 2015.
[12] C. Huang, R. J. Lamond, S. K. Pickus, Z. R. Li, and V. J. Sorger, "A sub-λ-size modulator beyond the efficiency-loss limit," *IEEE Photon. J.*, vol. 5, no. 4, p. 202411, Aug. 2013.
[13] Z. Ma, Z. Li, K. Liu, C. Ye and V. Sorger, "Indium-Tin-Oxide for High-performance Electro-optic Modulation", *Nanophotonics*, vol. 4, no. 1, 2015.
[14] D. Ansell, I. Radko, Z. Han, F. Rodriguez, S. Bozhevolnyi and A. Grigorenko, "Hybrid graphene plasmonic waveguide modulators", *Nature Communications*, vol. 6, p. 8846, 2015.
[15] E. M. Purcell "Spontaneous emission probabilities at radio frequencies" *Phys. Rev.* 69, 681, 1946
[16] A. Chaves, T. Low, P. Avouris, D. Çakır and F. Peeters, "Anisotropic exciton Stark shift in black phosphorus", *Phys. Rev. B*, vol. 91, no. 15, 2015.
[17] A. Chernikov, T. Berkelbach, H. Hill, A. Rigosi, Y. Li, O. Aslan, D. Reichman, M. Hybertsen and T. Heinz, "Exciton Binding Energy and Nonhydrogenic Rydberg Series in Monolayer WS 2", *Phys. Rev. Lett.*, vol. 113, no. 7, 2014.
[18] F. Wang, Y. Zhang, C. Tian, C. Girit, A. Zettl, M. Crommie and Y. Shen, "Gate-Variable Optical Transitions in Graphene", *Science*, vol. 320, no. 5873, pp. 206-209, 2008.
[19] Z. Lu and W. Zhao, "Nanoscale electro-optic modulators based on graphene-slot waveguides", *Journal of the Optical Society of America B*, vol. 29, no. 6, p. 1490, 2012.
[20] R. Hao, W. Du, H. Chen, X. Jin, L. Yang and E. Li, "Ultra-compact optical modulator by graphene induced electro-refraction effect", *Appl. Phys. Lett.*, vol. 103, no. 6, p. 061116, 2013.
[21] C. Phare, Y. Daniel Lee, J. Cardenas and M. Lipson, "Graphene electro-optic modulator with 30 GHz bandwidth", *Nature Photonics*, vol. 9, no. 8, pp. 511-514, 2015.
[22] W. Gao, J. Shu, C. Qiu and Q. Xu, "Excitation of Plasmonic Waves in Graphene by Guided-Mode Resonances", *ACS Nano*, vol. 6, no. 9, pp. 7806-7813, 2012.
[23] G. Hanson, "Dyadic Green's functions and guided surface waves for a surface conductivity model of graphene", *J. Appl. Phys.*, vol. 103, no. 6, p. 064302, 2008.
[24] V. Sorger, N. Lanzillotti-Kimura, R. Ma and X. Zhang, "Ultra-compact Silicon nanophotonic modulator with broadband response", *Nanophotonics*, vol. 1, no. 1, 2012.
[25] Q. Lin, O. Painter and G. Agrawal, "Nonlinear optical phenomena in Silicon waveguides: modeling and applications", *Opt. Express*, vol. 15, no. 25, p. 16604, 2007.
[26] A. Liu, R. Jones, L. Liao, D. Samara-Rubio, D. Rubin, O. Cohen, R. Nicolaescu and M. Paniccia, "A high-speed Silicon optical modulator based on a metal–oxide–semiconductor capacitor", *Nature*, vol. 427, no. 6975, pp. 615-618, 2004.
[27] Q. Xu, B. Schmidt, S. Pradhan and M. Lipson, "Micrometre-scale Silicon electro-optic modulator", *Nature*, vol. 435, no. 7040, pp. 325-327, 2005.
[28] G. T. Reed, G. Mashanovich, F. Y. Gardes, and D. J. Thomson, "Silicon optical modulators," *Nat. Photon.*, vol. 4, pp. 518–526, 2010.
[29] E. Feigenbaum, K. Diest and H. Atwater, "Unity-Order Index Change in Transparent Conducting Oxides at Visible Frequencies", *Nano Letters*, vol. 10, no. 6, pp. 2111-2116, 2010.
[30] L. Lafone, T. Sidiropoulos and R. Oulton, "Silicon-based metal-loaded plasmonic waveguides for low-loss nanofocusing", *Optics Letters*, vol. 39, no. 15, p. 4356, 2014.
[31] M. Liu, X. Yin and X. Zhang, "Double-Layer Graphene Optical Modulator", *Nano Letters*, vol. 12, no. 3, pp. 1482-1485, 2012.
[32] W. Barnes, "Surface plasmon–polariton length scales: a route to sub-wavelength optics", *Journal of Optics A: Pure and Applied Optics*, vol. 8, no. 4, pp. S87-S93, 2006.
[33] Narayana, V. K., Sun, S., Badawy, A. H. A., Sorger, V. J., & El-Ghazawi, T. (2017). MorphoNoC: Exploring the design space of a configurable hybrid NoC using nanophotonics. *Microprocessors and Microsystems*, 50, 113-126.
[34] Heidari, E., Dalir, H., Koushyar, F. M., Nouri, B. M., Patil, C., Miscuglio, M., ... & Sorger, V. J. (2022). Integrated ultra-high-performance graphene optical modulator. *Nanophotonics*, 11(17), 4011-4016.
[35] Ma, Z., Kikunaga, K., Wang, H., Sun, S., Amin, R., Maiti, R., et al. (2020). Compact graphene plasmonic slot photodetector on silicon-on-insulator with high responsivity. *ACS Photonics*, 7(4), 932-940.
[36] Amin, R., Tahersima, M. H., Ma, Z., Suer, C., Liu, K., Dalir, H., & Sorger, V. J. (2018). Low-loss tunable 1D ITO-slot photonic crystal nanobeam cavity. *Journal of Optics*, 20(5), 054003.
[37] Wang, H., Patil, C., Dalir, H., & Sorger, V. J. (2022, March). 20Gbps high-gain BW-product TMD slot-detector on PIC. In *2D Photonic Materials and Devices V* (p. PC120030A). SPIE.
[38] M. Nielsen, L. Lafone, A. Rakovich, T. Sidiropoulos, M. Rahmani, S. Maier and R. Oulton, "Adiabatic Nanofocusing in Hybrid Gap Plasmon Waveguides on the Silicon-on-Insulator Platform", *Nano Letters*, vol. 16, no. 2, pp. 1410-1414, 2016.
[39] Ye, C., Khan, S., Li, Z. R., Simsek, E., & Sorger, V. J. (2014). λ-size ITO and graphene-based electro-optic modulators on SOI. *IEEE Journal of Selected Topics in Quantum Electronics*, 20(4), 40-49.
[40] Amin, R., Ma, Z., Maiti, R., Khan, S., Khurgin, J. B., Dalir, H., & Sorger, V. J. (2018). Attojoule-efficient graphene optical modulators. *Applied Optics*, 57(18), D130-D140.
[41] Heidari, E., Dalir, H., Koushyar, F. M., Nouri, B. M., Patil, C., Miscuglio, M., ... & Sorger, V. J. (2022). Integrated ultra-high-performance graphene optical modulator. *Nanophotonics*, 11(17), 4011-4016.
[42] Hemnani, R. A., Tischler, J. P., Carfano, C., Maiti, R., Tahersima, M. H., Bartels, L., ... & Sorger, V. J. (2018). 2D material printer: a deterministic cross contamination-free transfer method for atomically layered materials. *2D Materials*, 6(1), 015006.
[43] Patil, C., Dalir, H., Kang, J. H., Davydov, A., Wong, C. W., & Sorger, V. J. (2022). Highly accurate, reliable, and non-contaminating two-dimensional material transfer system. *Applied Physics Reviews*, 9(1), 011419.
[44] Xu, Q., Schmidt, B., Pradhan, S., & Lipson, M. (2005). Micrometre-scale silicon electro-optic modulator. *nature*, 435(7040), 325-327.
[45] Tahersima, M.H., Ma, Z., Gui, Y., Sun, S., Wang, H., Amin, R., Dalir, H., Chen, R., Miscuglio, M. and Sorger, V.J., Coupling-enhanced dual ITO layer electro-absorption modulator in silicon photonics. *Nanophotonics*, 2019 8(9), pp.1559-1566.
[46] Amin, R., Maiti, R., Gui, Y., Suer, C., Miscuglio, M., Heidari, E., Khurgin, J.B., Chen, R.T., Dalir, H. and Sorger, V.J., Heterogeneously integrated ITO plasmonic Mach–Zehnder interferometric modulator on SOI. *Scientific reports*, 2021, 11(1), p.1287.
[47] Amin, R., Suer, C., Ma, Z., Sarpkaya, I., Khurgin, J.B., Agarwal, R. and Sorger, V.J., A deterministic guide for material and mode dependence of on-chip electro-optic modulator performance. *Solid-State Electronics*, 2017, 136, pp.92-101.
[48] Sorger, V.J., Amin, R., Khurgin, J.B., Ma, Z., Dalir, H. and Khan, S., Scaling vectors of attoJoule per bit modulators. *Journal of Optics*, 2017, 20(1), p.014012.
[49] Sorger, V. J., & Maiti, R. (2020). Roadmap for gain-bandwidth-product enhanced photodetectors: opinion. *Optical Materials Express*, 10(9), 2192-2200.
[50] V. J. Sorger, R. F. Oulton, R. M. Ma, X. Zhang, "Toward integrated plasmonic circuits" MRS bulletin 37 (08), 728-738, 2012.
[51] A. Fratalocchi, C.M. Dodson, R. Zia, P. Genevet, E. Verhagen, H. Altug, V. J. Sorger, "Nano-Optics gets practical" Nature Nanotechnology 10 (1), 11-15, 2015.
[52] Narayana, V.K., Sun, S., Badawy, A.H.A., Sorger, V.J. and El-Ghazawi, T., 2017. MorphoNoC: Exploring the design space of a configurable hybrid NoC using nanophotonics. *Microprocessors and Microsystems*, 50, pp.113-126.
[53] Miscuglio, M., Adam, G.C., Kuzum, D. and Sorger, V.J., 2019. Roadmap on material-function mapping for photonic-electronic hybrid neural networks. *APL Materials*, 7(10), p.100903.
[54] Miscuglio, M., & Sorger, V. J. (2020). Photonic tensor cores for machine learning. *Applied Physics Reviews*, 7(3), 031404.